\newcommand{\enq}{\end{quote}}
\newcommand{\be}{\begin{equation}}
\newcommand{\en}{\end{equation}}
\newcommand{\del}{\delta}

\documentclass[11pt]{amsart}
\usepackage{geometry}                
\usepackage{graphicx}
\usepackage{amssymb}
\usepackage{epstopdf}
\title{Atmospheric Refraction  Based on Atmospheric Pressure and Temperature Data}
\author{Michael Nauenberg\\ 
Physics Dept., University of Californa\\
Santa Cruz, CA. 95064
}
\begin{document}
\maketitle

\begin{abstract}
Calculations of atmospheric refraction are generally based on a simplified  model of atmospheric density in the troposphere which assumes that the temperature decreases at a constant lapse rate $L$ from sea level up to a height $h_t\approx 11$ km, and  that  afterwards  it  remains  constant.  In this model,  the ratio $T_o/L$,  where $T_o$ is  the temperature at the observer's  location,   determines the length scale in the calculations for altitudes $h\leq h_t$.
But daily balloon measurements across the U.S.A. reveal that  in some cases  the air temperature actually
increases from sea level up to a height $h_p$ of about one km,  and  only after reaching a  plateau with  temperature $T_o'$ at this height, it decreases at an {\it approximately}  constant  lapse rate.Hence, in such cases, the relevant length scale for atmospheric refraction 
calculations in the   altitude  range $h_p\leq h<h_t$  is  $T_o'/L$, and the contribution  for $h\leq h_p$ has to  be calculated from  actual measurements  of air density in this range. Moreover, in three examples considered here,  the temperature does not remain constant for  $h_t \leq h$, but  continues to decreases to  a minimum  at $h_m \approx 16$ km, and then increases at higher altitudes at a lower rate.  Calculations of  atmospheric refraction based on this atmospheric data  is  compared with the 
results of simplified  models.
\end{abstract}

\section*{Introduction}
In current models to  calculate  atmospheric refraction, it is assumed that  in the troposphere the temperature decreases from sea level at a  constant rate known as the  lapse rate  up to a height $h_t$, and that above  this height  the temperature remains essentially constant until reaching the stratosphere at $h >$ 20 km,  where it increases at a slow rate\footnote{  For example, in an influential paper, C.Y. Hohenkerk and A. T Sinclair state: ``The temperature decreases at a constant rate within the 
troposphere, up to the tropopause  at about  11 km height. Above the tropopause the temperature remains constant". They quote a lapse
rate $L\approx 6.5^0$ Celsius/km, but  their algorithm in current use  allows for different rates 
\cite{hohen}(see routine from SLALIB at http://star-www.rl.ac.uk/docs/sun67.htx/sun67ss157.html)}.  But  measurements by balloons released  daily into the atmosphere across the U.S.A. \cite{balloon} reveal  that this model is only approximately correct. In particular,  in some cases at an altitude below  about 1 km,
the temperature {\it increases}  instead of decreasing with altitude or decreases at a rate different from the asymptotic lapse rate.  Moreover, instead of a tropopause at $h_t\approx11$ km,   the temperature decreases to  a minimum at a  higher altitude, $h_t\approx 16$ km., and then increases at a slower rate\footnote{This effect is attributed to the occurrence of
an ozone layer }. 

To illustrate this behavior, three  examples of the air temperature dependence on altitude obtained by recent  balloon measurements are shown in Figs.1-4. Moreover, in these cases  the minimum temperature occurs at about $ 10^0$  Celsius below  the  temperature
 in the conventional model ($-57^0$ Celsius) , and the estimated lapse rates 
are different, ranging from $L=6^0$ to $8^0$  degrees  Celsius/km.\footnote{ In a spherically symmetric atmosphere,  one would  expect  that  it is not the lapse rate, but  the   minimum atmospheric temperature,  and the altitude  where it occurs in  the troposphere, that  is independent of the observer's location. This is found to be the case in the three examples discussed here, two of which are separated by a distance equal to  1/5 the Earth's circumference.}

For  sea level observations along the horizon,
the air density up to  an altitude of 1 km  contributes  about 1/4 of the total atmospheric refraction (an example is given in  Table I),  but  the effect  of a possible temperature increase up to this altitude has been ignored previously\footnote{ A detailed  discussion of  low- altitude refraction has been given by Andrew T. Young \cite{young}.}.  Instead, in calculations 
of atmospheric refraction it is  assumed that the temperature decreases uniformly at a constant lapse rate $L$ starting at the location
of the observations,  and consequently,  that  the temperature $T_o$ at this location determines the relevant length scale $T_o/L$ in the calculation  of atmospheric refraction. For these observations, this assumption  leads to significant errors in these calculations,  but of course it does not affect calculations  made for observations at altitudes above  about 1 km.  But the occurrence  of a minimum temperature instead of the assumed  constant value in the troposphere, leads to differences  in fractions of arc seconds. 
Our main conclusion  is that for  an accurate calculation of atmospheric refraction  at a given location and time,   the parameters of the  analytic expression for the atmospheric density should be based  on a fit  to the atmospheric data  obtained  at the nearest station.

In the next section the theory of  atmospheric  refraction for a spherically symmetric atmosphere is extended to the case that the linear decrease in temperature  in the troposphere is valid  only above a finite height $h_p$ . An  analytic approximation is derived for the contribution of atmospheric density below this altitude.  In section II,  we present the parameters
of the analytic  atmospheric model  discussed in section I, obtained  by a 
least square fit  to the  air  pressure and temperature  data at  three separate stations\footnote{Liheu, Kauai; Oakland,C.A; Buffalo, N.Y.} (see Table I). 
For a case when the  temperature increases with altitude at low altitudes, the results from  a fit to the  data at  the  Oakland station for $h \leq .6$ k are  compared with the standard calculation for $80^o\leq \phi \leq 90^o$, where $\phi$ is the angle of observation relative to the zenith of the observer (see Table II). For $4\leq h \leq 16$,  a fit to the data for the ratio pressure/temperature proportional to density,   is shown  in Fig.6. For an illustration of observations above $h \approx $ 1 km,  the refraction at the Keck observatory \footnote{This observatory is located on Mauna Kea at 4.207 km above sea level.} is calculated based  on the atmospheric data obtained by the  station at  Lihue, Kauai,   and compared with the  refraction table for this observatory (see Table III)

\begin{figure}[htbp] 
   \centering
   \includegraphics[width=11cm]{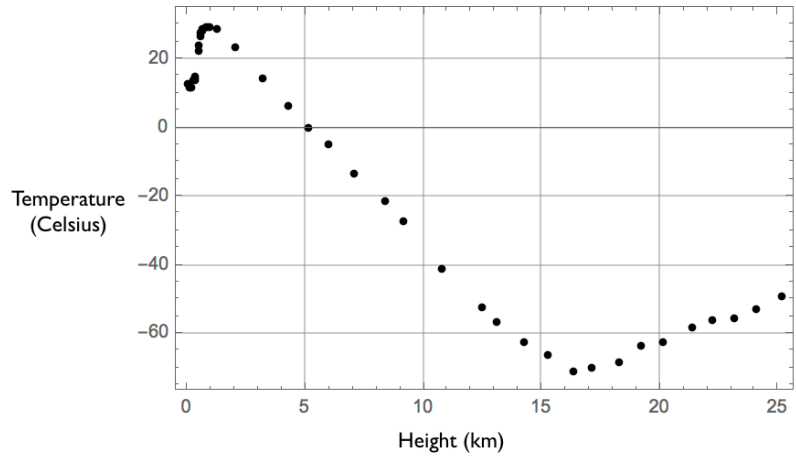}
\caption{Atmospheric temperature, Oakland, California, July 29, 2016.  }.
\label{  }
\end{figure}

\begin{figure}[htbp] 
   \centering
   \includegraphics[width=10cm]{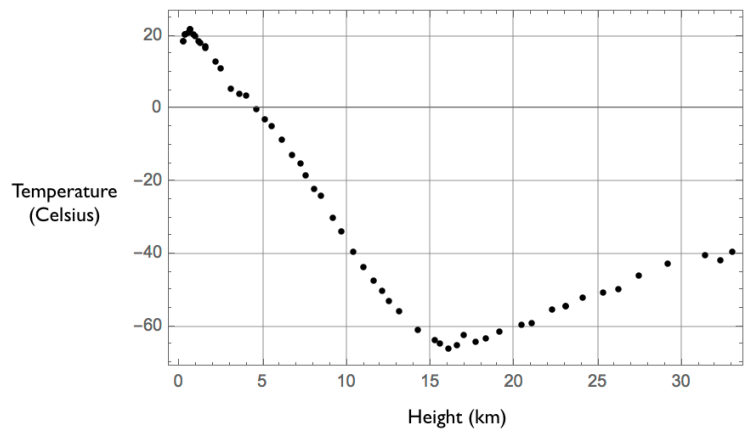}
\caption{Atmospheric temperature, Buffalo, N.Y., August 4, 2016.}.
\label{  } 
\end{figure}

\begin{figure}[htbp] 
   \centering
   \includegraphics[width=13
   cm]{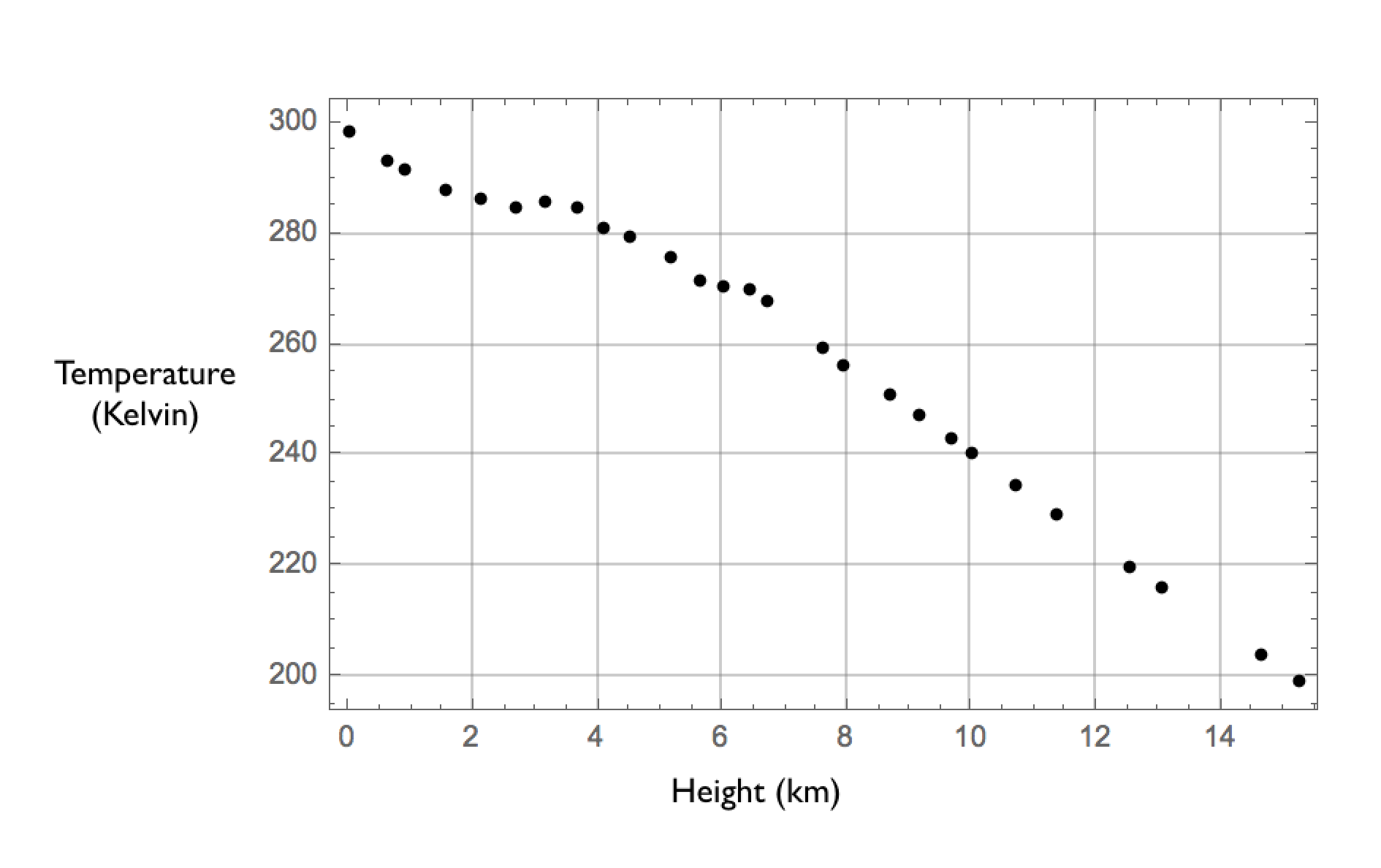}
\caption{Atmospheric temperature, Lihue, Kauai, August 20, 2016.  }.
\label{  }
\end{figure}

\begin{figure}[h!]
\centering
\includegraphics[width=12cm]{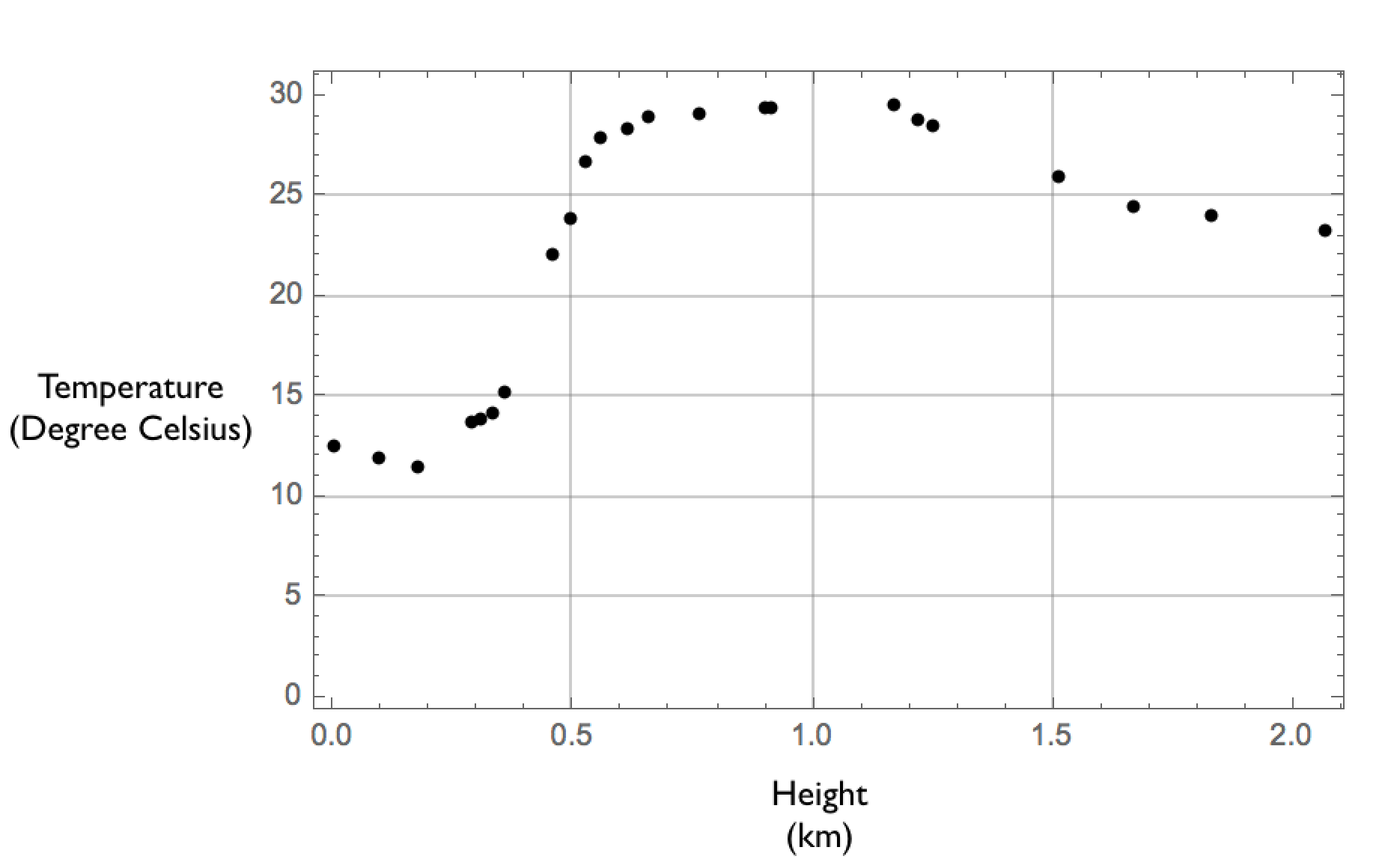}
\caption{  Low altitude atmospheric temperature, Oakland, July 29, 2016.}.
\label{}
\end{figure}

\clearpage

\section*{Theory}

The refraction of light  traversing a spherically symmetric  atmosphere is calculated by  the well known 
 integral
  \be
\label{refraction}
R(\phi)=-\int_{r_o}^{\infty}dr\: \frac{dn}{ndr}\frac{\sin\phi}{\sqrt{(nr/n_o r_o)^2-\sin^2\phi}}, 
\en
where $n(r)$ is the index of refraction of atmospheric air at a distance $r$ from the center of the Earth, $\phi $ is the  angle  relative  to the zenith for observations at 
 an altitude $h_o$ above sea level,  and $r_o=r_E+h_o$,  where $r_E$ is the mean radius of the Earth \cite{newcomb}.
  For $r \geq r_o$, $n( r)$  is assumed to be determined by the atmospheric air density $\rho(r)$ according to the relation

\be
\label{index}
n(r)=1+c(\lambda) \frac{\rho(r)}{\rho(0)}, 
\en
where  $c(\lambda)$ is a constant that depends on the wavelength $\lambda$ of the light beam \cite{ciddor},\cite{index}.\\

Since the density of atmospheric air is very low and decreases with altitude, it is reasonable to  assume that it
 satisfies the ideal gas relation,
\be
\label{ideal}
\rho=\frac{m}{k}\frac{p}{T},
\en
where $k$ is Boltzmann's constant,  $m$ is the mean particle mass of atmospheric air, $p$ is the pressure,  and $T$ is the temperature.
Assuming that  parcels of air at $r$ are in static equilibrium,  
\be
\frac{dp}{dr}=-g \rho,
\en
where $g$ is the gravitational acceleration at $r_0$ \footnote{The dependence of $g$ on altitude can be neglected for the altitudes considered  here.}.
Applying the ideal gas law, Eq.\ref {ideal}, we have  
\be
\label{pressure}
p(h)=p(h_o) exp( -\frac{mg}{k} \int_{h_o}^h  \frac{dh'}{ T(h')}),
\en
where $h=r-r_E$ .
Introducing the conventional  assumption that  for $h\geq h_o$ the temperature decreases linearly with  altitude \cite{ivory},\cite{garfinkel}
\be
\label{temperature}
T(h)=T_o -L ( h-h_o),
\en 
where  $L =-dT/dh$ is a constant known as the  lapse rate, and $T_o=T(h_o)$ is the temperature at the initial altitude $h_o$. Substituting
this relation  in Eq.\ref{pressure} for the pressure,  one  obtains
\be
\label{pressureb}
p( h)=p(h_o) (1 -\frac{(h-h_o)}{H_o})^q, 
\en
where the exponent $q$, given by 
\be
\label{qvalue}
q=\frac{mg}{kL}, 
\en
depends only  on  the lapse rate $L$, while  the length parameter 
\be
\label{lapserate}
H_o=\frac{T_o}{L},
\en
depends on both the lapse rate $L$, and on the temperature $T_o$ at the onset of the  assumed linear temperature decrease with altitude\footnote{ Since $T_o=T(0)-h_oL$, Eq.\ref{temperature}, $H_o=T(0)/L-h_o$}.  
Hence,  according to the ideal gas law, Eq.\ref{ideal},  for $h\geq h_o$ the air density  as a function of altitude is given by \cite{ivory},\cite{garfinkel}
\be
\label{density}
\rho(h)=p(h)\frac{T_o}{T(h)}= \rho(h_o)(1-\frac{(h-h_o)}{H_o})^{q-1},
\en
and the dependence of the index of refraction on   the altitude $h$, Eq.\ref{index}, is 
\be
\label{index2}
n(h)=1+c '(\lambda)(1-\frac{(h-h_o)}{H_o})^{q-1},
\en
where 
\be
c'(\lambda)= c(\lambda) \frac{\rho(h_o)}{\rho(0)}. 
\en

For  $h<h_p$,  where $h_p$ is approximately one  kilometer, the  assumption that the temperature decreases with a lapse rate $L$
is not generally valid, and in some cases balloon measurements \footnote{
These balloons are configured  to  measure the altitude dependence of air  pressure and temperature,  while the  air  density is  obtained from the ratio of these quantities, in accordance with the ideal gas law, Eq. \ref{ideal}.}  indicate that the temperaturet actually increases with altitude from sea level (see Figs 1-4). In this case, the air density can be approximated by a power series in h. To second order,
\be
\label{power}
\frac{\rho(h)}{\rho(0)}\approx 1+ah+bh^2, 
\en 
 and the  integration for the contribution $\del R(h,\phi)$ to the refraction integral, Eq.\ref{refraction}, in the interval $0\leq h'\leq h$, can be carried out analytically,  
\be
\label{refapp}
\del R(h,\phi)=c(\lambda) \sin\phi(A(a,b,h,\phi)+B(b,h,\phi)),
\en
where 
\be 
A(a,b,h,\phi)=(ar_o-br_o^2 \cos^2 \phi)(\sqrt{2(\frac{h}{r_o})+\cos\phi^2}-\cos\phi),
\en
and
\be
B(b,h,\phi)=\frac{br_o^2}{3}((\frac{2h}{r_o}+\cos^2\phi)^{3/2}-\cos^3\phi)).
\en
For illustration, in the case that  $\rho(h)$ is given by the the canonical expression, Eq.\ref{density},  the coefficients  are $a=(q-1)/H_o$,  and $b=(q-1)(q-2)/2H^2_o$. For $h<<H_o$ this  approximation for $\del R(h,\phi)$ is in good agreement with the  exact 
contribution from the  integral, Eq.\ref{refraction}.

Actually,  even at altitudes  higher than $h_t$,  the temperature decreases linearly only approximately, and therefore  Eqs. \ref{temperature} and \ref{pressureb} for the temperature and pressure are  also satisfied  only approximately,  as will be shown in the next section. In this case,   the exponent $q$ and the 
parameter $H_o$  should be obtained  by a least square fit of the data for the ratio {\it pressure/temperature} to the theoretical relation, Eq. \ref{density},  for the density.

\section*{Numerical calculations}
The parameters for the atmospheric model discussed in the previous section were obtained  by a least square fit to the atmospheric  data
in the range $1-16$ km  at three separate stations on a specific day \footnote{ Liheu, Kauai on August 20,  Oakland,C.A on July 29, and Buffalo, N.Y on August 4.}, and the results are shown in Table I.   For each column associated with one of these  stations, the lapse rate $L$  and the  temperature $T_o$ in the first and second column were obtained from a fit to the temperature data, while the exponent $q$, and the scale height  $H_o$  in the third and fourth row, associated with the pressure dependence on altitude, Eq.\ref{pressureb}, were obtained
from a fit to the    pressure/temperature ratio vs altitude. The  values of $T_o$ and $H_o$ correspond to values at  sea level, $h=0$ km, extrapolated from the fit to the data for $h\geq$ 1 km. The fifth row is the minimum temperature $T_m$  of the troposphere at the respective locations. For comparison, the last column shows the conventional values of these  parameters.  

 If the temperature decrease with altitude where truly linear,  the theoretical relations  for the lapse rate, $L=T_o/H_o$, Eq.\ref{lapserate}, and  for the exponent in the relation for the  pressure, $q=mg/kL$, Eq.\ref{qvalue},  should be satisfied exactly, but this is not actually the case.
In practice, these two relations are  satisfied  only approximately, reflecting the fact that the decrease in temperature is  only approximately linear. Since the  variable  that is  relevant for  the calculation of atmospheric refraction, Eq.\ref{}, is the  atmospheric density $\rho$,  which
according to the ideal gas law,  is
proportional to the ratio  pressure/temperature, Eq.\ref{ideal},  the most appropriate  parameters for this calculation   should be the values  for  $q$ and $H_o$  obtained by a least square fit of the theoretical  density dependence on altitude, Eq.\ref{density},  to the observed altitude dependence of this ratio. 

Above the height $h\approx 16 $ km,  where  the  minimum temperature  $T_m$ occurs,  the atmospheric  temperature increases with altitude at a slow rate, but the balloon data is not always available at these higher altitudes.  In this case the best  approximation  is to assume that the temperature remains constant, which implies that the density decreases exponentially with a length scale $ H_m=kT_m/mg$.

To illustrate the contribution  to atmospheric refraction of air density at  very low altitudes (less than about  one kilometer), in a  case that the
temperature  increased  with altitude, consider the data  obtained by  balloon measurements  at  Oakland, CA.,  on July 29, 2016 \cite{balloon}.  Below an altitude of about $.6$ km, the temperature increased  from about $13^o$ to $28^o$ Celsius,  where  it reached a plateau (see Fig.4)  before decreasing  approximately linearly to  a minimum at about $16$ km (see Fig.1 ).  
The dependence of the  ratio  pressure/temperature on altitude,  is shown in Fig.5.   A least square fit of this ratio to a power series  to second order in $h$, Eq.\ref{power},  gives  $a=-.034 $ and b=-.309, is also shown in this figure. The  contribution $\del R(h,\phi)$ to the refraction,
Eq.\ref{refapp}, in the range $80^o\leq \phi \leq 90$  is given in Table I, and compared to the contribution calculated with the conventional model.   For higher altitudes the  pressure/density  data  is shown in Fig. 6,   with a fit based on  Eq. \ref{density}  with  the parameters
describe above. 

In Table II, a calculation of the atmospheric  refraction at the Mauna Kea observatory,  for $\lambda=.633$ microns,  based on
a least square fit to the atmospheric pressure and temperature measurements  at Lihue, Kauai, is compared with a current  refraction table  at this  observatory\footnote{ I would like to thank  Drew Phillips for a copy of the refraction  table at the Mauna Kea telescope}.

\begin{figure}[htbp] 
   \centering
   \includegraphics[width=15cm]{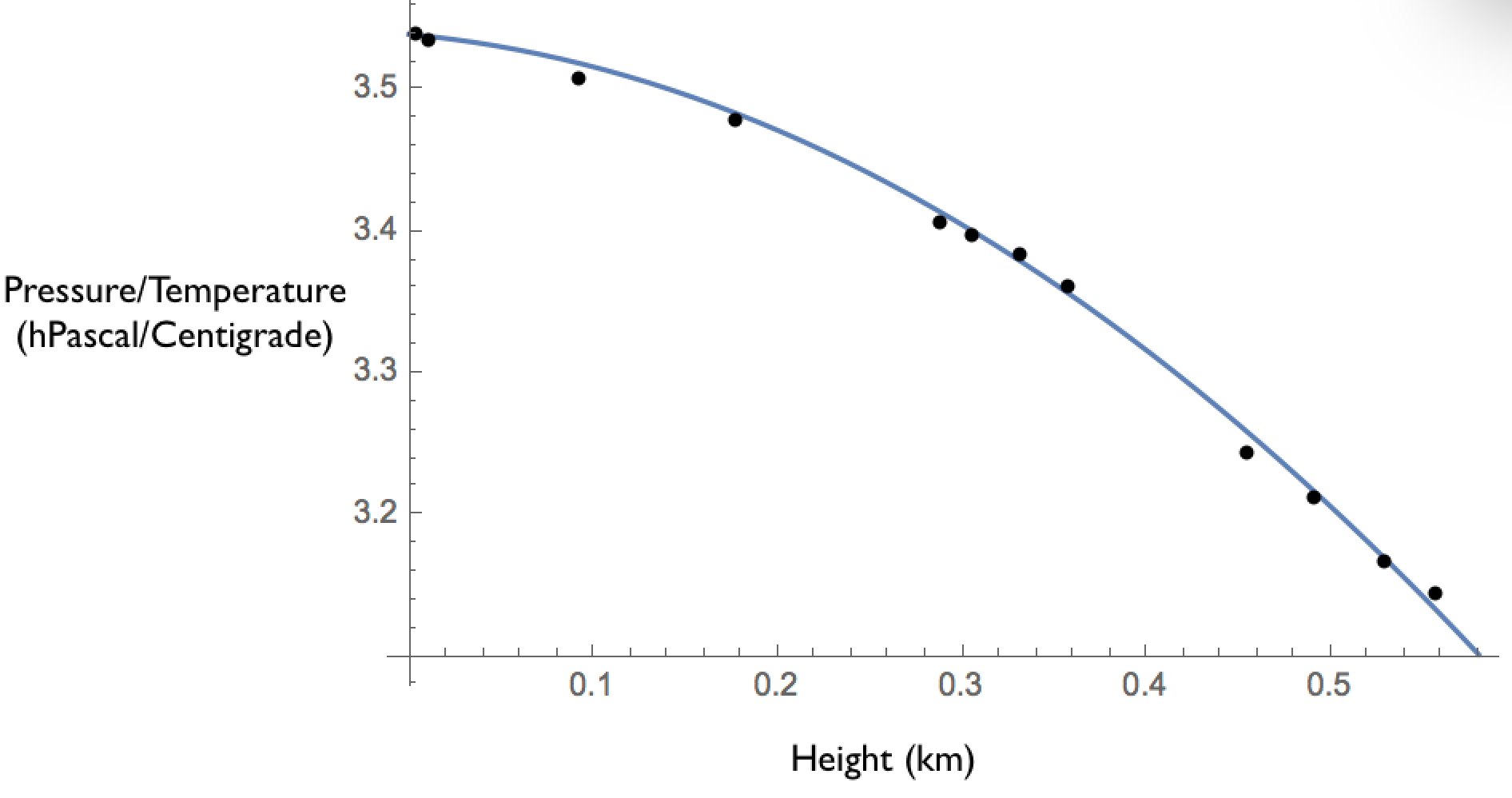}
   \caption{Low altitude pressure/temperature data, Oakland, CA. July 29, 2016. }
\label{  }
\end{figure}

\section*{ Table I}
\section*{Refraction parameters  obtained from a least square fit to balloon measurements of  atmospheric pressure and temperature at the
locations shown. The last column  are the parameters in conventional calculations currently in use.}

\begin{tabular}{ |p{3cm}||p{2cm}|p{2cm}|p{2cm}|p{2cm}|}\hline
 \multicolumn{5}{|c|}{ {\bf Refraction Parameters from fit to Troposphere data}} \\ \hline
 Parameters &Lihue & Oakland&Buffalo&Conventional\\ \hline
$T_o$ (Kelvin)& 319.3& 310.9&307&288.15\\ \hline
 L(Celsius/km) & 7.91&7.19&6.95&6.5\\ \hline  
q(exponent)  &4.31&4.59&5.16&5.25 \\ \hline
$H_o$ (km)  &40.35&41.71&46.3
&44.33\\ \hline
$T_m$ (Kelvin)  &197&203&208&216\\ \hline
\end{tabular}

\section*{Table II}
\section*{ The second column  is based on the conventional assumption that the
 lapse rate is  $L=6.5^0$ degrees Celsius/km. The third column is based on a fit to actual  atmospheric data obtained at the Oakland station on July 29, 2016.
 }

\begin{tabular}{ |p{2cm}||p{2cm}|p{4cm}|}
 \hline
 \multicolumn{3}{|c|}{ \bf Contribution to Atmospheric Refraction  } \\ \hline
  \multicolumn{3}{|c|}{ \bf (1-.6 km altitude), Oakland, CA. July 29.2016 }\\ \hline
Angle from Zenith in degrees &Conventional calculation. arc minutes& Least square fit to pressure/temperature data.  arc minutes 
 \\ \hline
 \hline
 $80^o$ & .31&.71\\ \hline  
 $81^o$  &.34&.79  \\ \hline
 $82^o$  &.38&.89\\ \hline
 $83^o $& .44& 1.02\\ \hline
 $84^o $&.51&1.18\\ \hline
 $85^o$  & .62& 1.42\\ \hline
 $86^o$  & .77& 1.76\\ \hline
 $87^o$  &1.02&2.33\\ \hline 
 $88^o$  & 1.50&3.38\\ \hline 
 $89^o$  &2.77&5.88\\ \hline 
 $90^o$  & 8.52&11.82\\ \hline 
 \hline
\end{tabular}

\begin{figure}[h!]
\centering
\includegraphics[width=15cm]{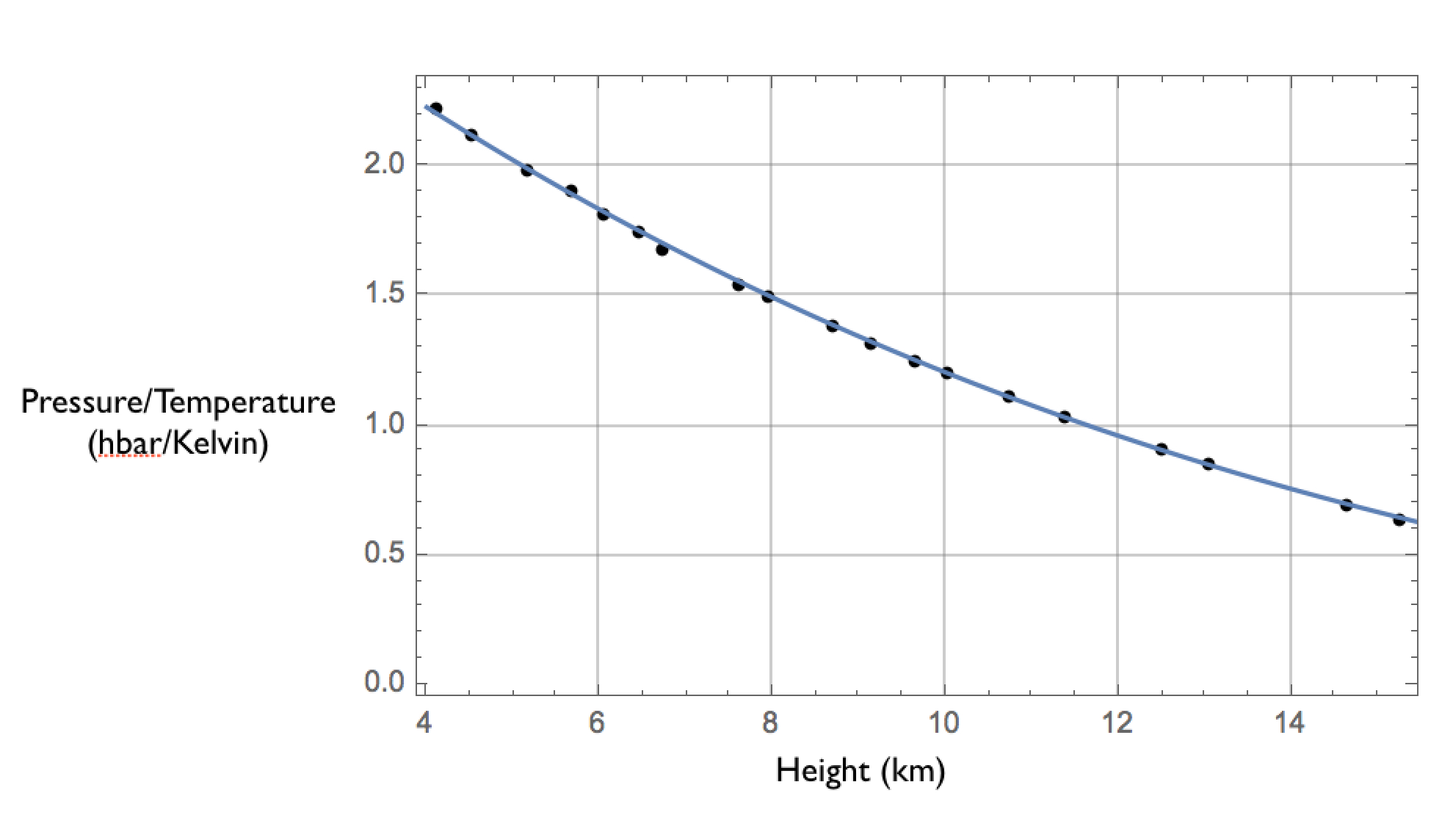}
\caption{Pressure/ Temperature vs. altitude data, Oakland, July 29, 2016. The curve is a fit based on Eq.\ref{density}, with the
exponent  parameter
$q=4.59$, corresponding to a lapse rate $L=7.19^0$ degrees Celsius/km.}.\
\label{}
\end{figure}

\section*{Table III}
\section*{Table  of atmospheric refraction at the Mauna Kea observatory, for  $\lambda=.633$ microns. The first column are observations angles relative to the observer's zenith.  The second column is  a refraction  table at the Mauna Kea observatory. The third column was calculated from the conventional parameters. The fourth column was  calculated from a least squares fit to  the Lihue data for the ratio
pressure/temperature dependence on altitude.
in Kauai,  on August 20,2016.}
 
\begin{tabular}{ |p{2cm}||p{2cm}|p{2cm}|p{2cm}|}
 \hline
 \multicolumn{4}{|c|}{ \bf Atmospheric Refraction at  Mauna Kea} \\ \hline
Zenith angle,  degrees & Mauna Kea Table, arc  seconds  & Conventional theory,
 arc  seconds&Lihue data, arc seconds  \\ \hline
 \hline
  $45^o$ & 36.70&36.80&36.81\\ \hline  
 $50^o$ & 43.72&43.83&43.84\\ \hline  
 $55^o$  & 52.35 & 52.47&52.50\\ \hline
 $60^o$  & 63.41 & 63.54&63.59\\ \hline
 $65^o $& 78.35 & 78.48&78.56\\ \hline
 $70^o $&100.0 &100.1&100.2\\ \hline
 $75^o$  & 134.7 &134.7&135.0\\ \hline
 $80^o$  & 200.0 & 199.6&200.6\\ \hline
 $85^o$  & 361.8 &361.6&366.5\\ \hline

 \hline
\end{tabular}

\section*{Summary}
Daily balloon measurement of atmospheric air pressure  and temperature show that the conventional assumption that the temperature in the troposphere decreases linearly with temperature up to
an altitude $h=11$ km,  followed by a constant temperature layer up to $h=20$ km, the so-called tropopause,  is not  generally valid.  For illustration,  plots of 
the altitude dependence of  the  temperature data from three separated  atmospheric stations, taken at  different dates,  are shown in Figs.1-4. This data shows  that in some cases at low altitudes, the temperature actually increases with altitude up to a height $h\approx$ 1 km, and that  instead of a tropopause, the temperature continues to decrease to a minimum value at $h\approx$ 16 km,  and afterwards it  increases  slowly at higher altitudes. To calculate the contribution of low altitude air density  to  the refraction integral,  the atmospheric data in the range $h$ from  0 to approximately 1km can be fitted  by a an expansion in 
powers of $h$. In the  range $1 \leq h \leq 16 $, 
a  least square fit to the atmospheric temperature data assuming  a linear decrease in temperature with height, Eq.\ref{temperature}, determines the lapse rate $L$, while a corresponding fit to the data on the ratio of pressure to temperature by the theoretical  relation for the air density $\rho$, 
Eq.\ref{density}, determines the exponent $q$.  According to the theory,   the mean molecular mass of the atmospheric air is $m=qLk/g$, and
substituting the values  obtained from the Liheu data, Table I, one obtains $m$= 28.96 gram/mole,  in remarkably  good agreement with the value  from measurements of air density at sea level. But  for the fit to the  data from the Oakland and Buffalo stations (see Table i) this agreement is only approximate, reflecting the fact that the  temperature decrease on these locations was  only approximately linear.

Finally, it should be pointed out that although calculations of atmospheric refraction based on simplified models  are usually given to seven or
more significant figures \cite{pulko},  this precision  is  meaningless. Accuracy in calculations of atmospheric refraction to more than seconds of arc  is not attainable without  fitting the parameters of  the analytic model for  the dependence of  the atmospheric density with height  to  
actual data.

\section*{Acknowledgements}

I would like to thank  G. Albrecht,  M. Berry, E. Falco, R. Kibrick, and S.Vogt for helpful comments, and C. Bohren, P. Chuang,  D. Phillips,  A. Klemola and A.T. Young  for references.

\end{document}